\documentclass[useAMS,usenatbib]{mn2e}
\usepackage{graphicx}
\def\rmscr#1{{\hbox{\rm \scriptsize #1}}}
\def\rmmat#1{{\hbox{\rm #1}}}
\def\d{\rmmat{d}}

\def\msun{{\rm M}_\odot}
\begin{document}
\title[White-Dwarf Kicks III.]{Constraining white-dwarf kicks in globular
  clusters : III.~Cluster Heating}

\author[J. Heyl]{Jeremy Heyl$^{1}$\\
$^{1}$Department of Physics and Astronomy, University of British Columbia, Vancouver, British Columbia, Canada, V6T 1Z1 \\
Email: heyl@phas.ubc.ca; Canada Research Chair}

\date{\today}

\pagerange{\pageref{firstpage}--\pageref{lastpage}} \pubyear{2007}

\maketitle

\label{firstpage}

\begin{abstract}
  Recent observations of white dwarfs in globular clusters indicate
  that these stars may get a velocity kick during their time as
  giants.  This velocity kick could originate naturally if the mass
  loss while on the asymptotic giant branch is slightly asymmetric.
  The kicks may be large enough to dramatically change the radial
  distribution of young white dwarfs, giving them larger energies than
  other stars in the cluster.  As these energetic white dwarfs travel
  through the cluster they can impart their excess energy on the other
  stars in the cluster.  This new heat source for globular clusters is
  expected to be largest during the clusters' youth.
\end{abstract}

\section{Introduction}
\label{sec:introduction}

\citet{1998A&A...333..603S} proposed that white dwarfs can acquire
their observed rotation rates from mild kicks generated by asymmetric
and off-centered  winds toward the end of their time on the asymptotic
giant branch (AGB) \citep{1993ApJ...413..641V}.
\citet{2003ApJ...595L..53F} invoked these mild kicks to explain a
putative dearth of white dwarfs in open clusters
\citep[e.g.][]{1977A&A....59..411W,2001AJ....122.3239K}.
\citet{2008MNRAS.383L..20D} observed a possible signature of
white-dwarf kicks in NGC~6397, and \citet{2007arXiv0712.0602C} found
similar but weaker hints in Omega
Centauri. \citet{2008MNRAS.383L..20D} found that young white dwarfs
are less centrally concentrated than either their progenitors near the
top of the main sequence or older white dwarfs whose velocity
distribution has had a chance to relax.

Early in the life of a cluster mass loss due to stellar evolution
competes with the loss of stars due to evaporation from the cluster
\citep{Spit87}.  Without a kick young white dwarfs would have a
velocity distribution nearly equal to that of their more massive
progenitors on the main sequence.  In this case the kinetic energy of
these white dwarfs is much less than equipartition; therefore, as
their velocity distribution relaxes they cool the rest of the cluster.
If on the other hand white dwarfs receive a substantial kick at birth
as observations indicate \citep{2008MNRAS.383L..20D}, young white dwarfs
may heat the rest of the stars in the cluster.  This letter will examine
how white dwarf kicks affect the energy balance within a globular
cluster.

\section{Calculations}
\label{sec:calculations}

Clusters of stars can typically be modelled with a lowered isothermal
profile (or King model) \citep{1963MNRAS.125..127M,1966AJ.....71...64K,Binn87}.
\begin{equation}
f = \frac{\d N}{\d^3 x \d^3 p} = 
 \left \{ 
\begin{array}{ll}
\rho_1(2\pi\sigma^2)^{-3/2} \left ( e^{\epsilon/\sigma^2}- 1 \right )
&  \rmmat{~if~} \epsilon>0 \\
0 & \rmmat{~if~} \epsilon\leq 0 
\end{array}
\right .
\label{eq:1}
\end{equation}
where $\epsilon = \Psi - \frac{1}{2} v^2$ and $\Psi$ is the
gravitational potential. Because the distribution function depends only
on constants of the motion (the energy), it is constant in time as
well.

With time the kinetic energy within the cluster approaches
equipartition between the various stars such that $m_i \sigma_i^2 =
m_j \sigma_j^2$ \citep{Spit87}.  The progenitors of young white dwarfs
will be the most massive main-sequence stars in a cluster at the time,
so they will typically have
$\sigma_\rmscr{TO}<\sigma_\rmscr{cluster}$, where
$\sigma_\rmscr{cluster}$ is the mean velocity dispersion of the
cluster.  As these stars evolve they lose mass.  During a globular
cluster's youth, this stellar mass loss dominates the mass loss from
the cluster; later the relaxation of the stellar velocity
distribution to a Maxwellian and the subsequent evaporation of stars
from the cluster dominates.  

\subsection{Kicks}

\begin{figure}
 \includegraphics[width=3.4in]{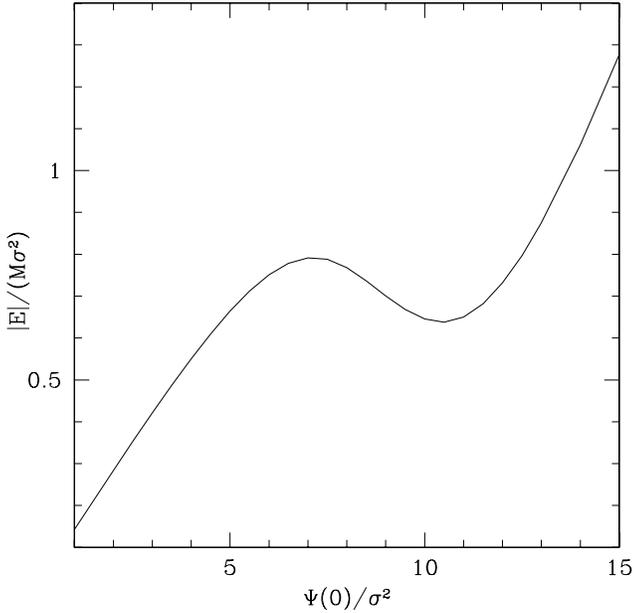}
 \caption{The total binding energy of a cluster modelled with a
   lowered isothermal distribution function as a function of the
   central gravitational potential.  For small $\Phi(0)$, the total
   binding energy increases linearly with the central potential.  For
   large values of $\Phi(0)$, $|E| \propto \Phi(0)^2$. }
 \label{fig:binding}
\end{figure}

At middle age the mass of white dwarfs that remain in the cluster may
exceed a third of the total mass of the cluster; consequently, if
white dwarfs receive a kick comparable to the velocity dispersion of
the cluster \citep{2008MNRAS.383L..20D,Heyl07kickgc,Heyl07kickobs},
the total kinetic energy of the kicks may approach 10-20\% of the
binding energy of the cluster as shown in Fig.~\ref{fig:binding}.

Specifically, \citet{Heyl07kickobs} found that an initial distribution
of white-dwarf progenitors with $\sigma_\rmscr{TO}=0.5\sigma$ and a
typical kick velocity of $\sigma_k=1.84\sigma_\rmscr{TO}$ could
explain the observations \citep{2008MNRAS.383L..20D}.  Essentially,
young white dwarfs receive a kick of the same order as the velocity
dispersion of the cluster. Furthermore, unlike neutron stars whose
kicks nearly always cause them to leave the cluster, most of the young
white dwarfs remain in the cluster to heat it up (only about two
percent escape within a crossing time).

These results provide an estimate of the total power in white-dwarf
kicks,
\begin{equation}
 \label{eq:kickpower}
  \frac{\epsilon_\rmscr{kick}\tau}{M \sigma^2} \approx
   \frac{\zeta}{2} \left | \frac{d\ln N}{d\ln m} \left ( \frac{d \ln \tau}{d \ln
         m} \right)^{-1}
  \right |_{m=m_\rmscr{TO}} \!\!\!\! \frac{0.38 \msun + 0.15
    m_\rmscr{TO}}{\bar m}
\end{equation}
where $dN/dm$ is the number of stars per unit mass in the cluster (the
mass function), $M$ is the total mass of stars in the cluster, $\bar
m$ is the mean mass of stars in the cluster, $\tau(m)$ is the duration
of the main sequence for a star in the cluster,
$\tau=\tau(m_\rmscr{TO})$ (the age of the cluster), and
$\zeta=\left(v_\rmscr{kick}/\sigma\right)^2\approx 0.85$.
\citet{1983ARA&A..21..271I} give the initial-final mass relation in
the numerator of the rightmost expression.

\citet{2007arXiv0708.4030R} estimate the current mass function for a
region near the half-light radius of NGC~6397 to be
\begin{equation}
  \label{eq:massfunction}
\frac{dN}{dm} = A m^{-\alpha} , \frac{d\ln N}{d\ln m} = 1-\alpha
\end{equation}
where the slope of the mass function today is given by $\alpha=0.13$
and $m_\rmscr{TO}\approx 0.8\msun$ ($\tau_0\approx 12$~Gyr is the
current age of the cluster) and $d\ln \tau/d\ln m\approx -3.75$
\citep{1991ApJS...76..525S} for a cluster of the age and metallicity
of NGC~6397.

Integrating Eq.~(\ref{eq:massfunction}) over the masses of the stars
in the cluster gives the mean mass of a star in the cluster (assuming
$\alpha<1$ and $m_\rmscr{TO} \gg m_\rmscr{min}$, the minimum mass of a
star),
\begin{equation}
  \label{eq:totalmass}
  \bar m = \frac{1-\alpha}{2-\alpha} m_\rmscr{TO} \approx 0.37 \msun
\end{equation}
so
\begin{equation}
  \label{eq:kickpower2}
  \frac{\epsilon_\rmscr{kick}\tau}{M \sigma^2} \approx
   \frac{\zeta}{2} \frac{ 2 - \alpha}{\beta}
\left ( \frac{0.38 \msun}{m_\rmscr{TO}} +  0.15 \right ) \approx 0.13
\end{equation}
for the current observations of NGC~6397.

\subsection{Binaries}

In a binary a fraction of the energy of the kick is used to change the
orbital parameters \citep{Heyl07kickbin}.  If the masses of the
primary and secondary are similar due to dynamical biasing
\citep[e.g.][]{1993MNRAS.262..800M}, the total kick to the binary is
about 70-80\% of the kick imparted to a single star, so given that the
fraction of binaries is small in the cluster as a whole and the
correction to the kick for binaries is also small, the energetics of
changing the orbits of binaries will be ignored.

On the other hand, the binaries provide an important energy source for
the cluster (in fact the only energy source if one excludes the
kicks).  The interaction of a binary with a single star can result in
an exchange or the dissolution of the binary, but generally it results
in an increase in the kinetic energy of final single star (like a
kick) at the expense of the increased binding energy of the binary.
The velocity increment of the singleton is generally random (also like
a kick), so these two energy sources are similar and possibly
comparably important to the evolution of the cluster.

\citet{Spit87} gives an estimate for the power from binaries
of 
\begin{equation}
  \label{eq:binary}
  \epsilon_\rmscr{binary} t_r \approx \frac{0.92}{\ln \Lambda} N_b
  \frac{\bar m \sigma^2}{2} 
\end{equation}
per relaxation time,
\begin{equation}
  \label{eq:relax}
  t_r \approx \frac{\sigma^3}{1.22 n_s 4\pi G^2 \bar m^2 \ln \Lambda}.
\end{equation}
where $N_b$ is the number of binaries. Combining these results yields
\begin{eqnarray}
  \label{eq:binary_power}
\frac{\epsilon_\rmscr{binary}\tau}{M\sigma^2} \!\!\!\!\!\! & \approx & \!\!\!\!\!\! 7.1 n_b \left
  (\frac{G \bar m}{\sigma} \right )^2 \sigma \tau \\
\!\!\!\!\!\!& \approx &\!\!\!\!\!\! 0.22  \left ( \frac{\bar m}{0.37 \msun}
\right )^2  \frac{n_b}{1 {\rm pc}^{-3}}
\left ( \frac{1~{\rm km s}^{-1}}{\sigma} \right )^3
\frac{\tau}{12~{\rm Gyr}}.
\end{eqnarray}
where the binary fraction is taken to be around a few percent suitable for
a typical region of the cluster outside the core \citep{2008arXiv0803.0005D}.
The power produced by the binaries is of course proportional to the
number density of the binaries and inversely proportional to the
velocity dispersion that sets the cross-section per binary.   

Comparing Eq.~(\ref{eq:binary_power}) to~(\ref{eq:kickpower2}) shows
that the power sources are similar for the region of NGC~6397 observed
by \citet{2007arXiv0708.4030R}.

\subsection{Evolution}

To look at the relative importance of binaries and kicks early in
the life of the cluster, some assumptions about the mass function of
the cluster in the past are needed.  First, the mass function becomes
more and more top heavy with time as the low mass stars are lost from
the cluster; therefore, it is natural to assume that $\alpha>1$ in the
past and possibly $\alpha>2$ near the turnoff.  In this regime, the
derivation of Eq.~(\ref{eq:kickpower2}) is not valid.  The
result in general is
\begin{equation}
  \label{eq:kickpower_old}
  \frac{\epsilon_\rmscr{kick}\tau}{M \sigma^2} \approx
   \frac{\zeta}{2} \frac{ |1 - \alpha|}{\beta}
\frac{0.38 \msun +  0.15 m_\rmscr{TO}}{\bar m} \sim \left (0.5 -
  1\right ) \zeta
\end{equation}
where the various slopes are evaluated at the turn-off.  Using the IMF
of \citet{1993MNRAS.262..545K} below one solar mass and
\citet{1986FCPh...11....1S} above gives $|1-\alpha|\approx 1.7$
compared to $2-\alpha \approx 1.9$ currently.  Also the value of the
$\beta$ only changes with time slightly (for $m_\rmscr{TO} \sim
1\msun$, $\beta\approx 3$), so the bulk of the increase comes from the
replacement of the turn-off mass with the mean mass in the denominator
of the expression.

Of course as the cluster evolves, the velocity dispersion of the
cluster should also evolve.  Because the mass of the cluster was
larger in the past, one would expect that the velocity dispersion was
also larger.  On the other hand, the kick velocity may also change
with the turn-off mass, so it is natural to introduce both of these
quantities as variables.  The power from binaries also depends
sensitively on the velocity dispersion and mean stellar mass.  Taking
the ratio of the kick power to the binary power yields,
\begin{eqnarray}
  \label{eq:ratio}
  \frac{\epsilon_\rmscr{kick}}{\epsilon_\rmscr{binary}} &=& 0.07
\frac{ |1 - \alpha|}{\beta} \frac{\left ( 0.38 \msun +  0.15
    m_\rmscr{TO} \right ) v_k^2 \sigma}{G^2 \bar m^3 n_b \tau } \\
&\approx& 40 \frac{v_k^2 \sigma}{(1~{\rm kms}^{-1})^3} \frac{1~{\rm
    Gyr}}{\tau} \frac{1~{\rm pc}^{-3}}{n_b}
\end{eqnarray}
where the approximation holds for $\tau = 10^7 - 10^9$~yr.  After
about a billion years, one would expect the cluster to have evolved
structurally, ejecting many of the low mass stars.  This would change
the mass function and typically decrease this ratio further.  The
increased velocity dispersion of the cluster actually increases the
relative importance of kicks early in the life of the cluster by
increasing the relaxation time (decreasing the binary power).  On the
other hand, the number density of binaries was likely to be larger in
the past than today simply because the number of density of stars was
larger then.

\section{Conclusions}
\label{sec:conclusions}
 
Over the life of a globular cluster such as NGC~6397, white-dwarfs
kicks may provide a significant energy source.  In a region outside
the core todayt, kicks provide about about one-half of energy input
from binaries.  Early in the life of the globular cluster when stars
of several solar masses are leaving the main sequence, the white-dwarf
kicks may actually dominate over binaries as an energy source;
consequently, young globular clusters such as those in starburst
galaxies may actually differ structurally from their older peers.
\citet{1992ApJ...386..106D} found that an energy source beyond
binaries was required to avoid core collapse in M71; perhaps white
dwarf kicks could explain this discrepancy. Regardless, these
calculations indicate that further study of the effects of white-dwarf
kicks on the dynamics of globular clusters is warranted.

\section*{Acknowledgments}

I would like to thank Harvey Richer and Saul Davis for useful
discussions and the referee, James Binney, would provided many useful
comments.  The Natural Sciences and Engineering Research Council of
Canada, Canadian Foundation for Innovation and the British Columbia
Knowledge Development Fund supported this work.  Correspondence and
requests for materials should be addressed to heyl@phas.ubc.ca.  This
research has made use of NASA's Astrophysics Data System Bibliographic
Services

\bibliographystyle{mn2e}
\bibliography{mine,wd,physics,math}
\label{lastpage}
\end{document}